\documentclass{pasj00}

\SetRunningHead{S.\ Hozumi}{Destructible Bars under
the Influence of a Massive BH}
\Received{2011/4/11}
\Accepted{2011/8/9}
\Published{2012}

\begin{document}
\title{Destructible Bars in Disk Galaxies\\ under the Dynamical
Influence of a Massive Central Black Hole}
 
\author{Shunsuke \textsc{Hozumi}}
\affil{Faculty of Education, Shiga University, 2-5-1 Hiratsu, Otsu,
Shiga 520-0862}
\email{hozumi@edu.shiga-u.ac.jp}
 
 
\maketitle

\begin{abstract}
The characteristics of the galactic bars that are prone to
suffer a damaging impact from a massive central black hole
are examined using flat stellar disks.  We construct three
disk model groups that consist of exponential disks with
one type of velocity distribution and Kuzmin-Toomre disks
with two different types of exact equilibrium distribution
function.  For each disk model group, three disks that
have different typical Toomre's $Q$ values are evolved
to form bars through dynamical instability.  Once a bar
is fully developed, a black hole (BH), whose mass is 1\%
of the disk mass, is adiabatically added at the center
of the disk.  Our results show that lower-amplitude bars,
that is, weaker bars are dissolved more easily by that BH.
We have found that this destructibility is rooted in the
characteristic feature that the bar formed spontaneously
becomes shorter in length and rounder in shape with
decreasing bar amplitude.  Since such weaker bars are
found to originate from colder disks in each disk model
group, it follows that for a given form of velocity structure,
the coldness of an initial disk determines whether the bar
produced in that disk is favorable to dissolution induced
by a massive central BH.  In addition, the existence of
bar-dissolved galaxies of the kind studied here is also
discussed.
\end{abstract}

\section{Introduction}
The existence of the supermassive black holes, whose masses
range from $\sim 10^6 M_\odot$ to $\sim 10^9 M_\odot$, that
reside in the centers of disk galaxies is gaining a firm
basis observationally (\cite{magorrian98,tremaine02,mh03,
marconi03,hr04,atkinson05,pastorini07,beifiori09,siopis09}).
Such a large-mass black hole (BH) could dissolve a bar
in a barred galaxy that embraces it, as a dynamical
consequence.  This kind of bar dissolution was first
studied by \citet{hn90}, who examined the stellar
orbits moving in combined fixed potentials of a rotating
bar and a central BH in two-dimensional configurations.
Next, \citet{hpn93} extended that work to three-dimensional
configurations.  Both studies showed that central mass
concentrations like massive central BHs can destroy a bar
by converting bar-supporting orbits into chaotic ones.

Although such orbital studies mentioned above are useful
for understanding the mechanism of bar destruction, they
cannot provide answers to the questions of how much BH mass
is needed, and how long it takes, precisely to destroy a bar.
As a helpful tool for resolving these issues, self-consistent
$N$-body simulations have played an important role in
investigating the detailed evolution of a bar under the
influence of a massive central BH.  As pioneering work on
$N$-body studies, \citet{nsh96} explored the evolution of
a Kuzmin-Toomre \mbox{(K-T)} disk \citep{kuzmin56,toomre63}
both in two- and three-dimensional configurations in which
a bar was formed by the bar instability, and then a BH was
added at the center of the disk as an external field.  They
showed that regardless of disk geometries, the minimum BH
mass required for bar destruction is $4-5$\% of the disk
mass.  Subsequently, \citet{ss04} have obtained a similar
BH mass to destroy a bar with high-quality $N$-body simulations
using a three-dimensional \mbox{K-T} disk embedded in a halo.

Unfortunately, K-T disks are inadequate in that real disk
galaxies are well-described by exponential surface-density
distributions \citep{freeman70}.  Then, \authorcite{hh05}
(\yearcite{hh05}, hereafter HH) adopted a two-dimensional
exponential disk and examined the effect of a central BH
on a bar generated in that disk.  Although the motions of
stars in their simulations are restricted to a single plane,
they have demonstrated that a BH with a mass of only 0.5\%
of the disk mass is sufficient for bar dissolution.  This BH
mass is about an order of magnitude smaller than that obtained
by Norman \etal\ (\yearcite{nsh96}) and that by \citet{ss04}.
Concerning this difference, HH argued that since exponential
disks are more centrally concentrated than K-T disks, central
BHs could have a more destructive impact on a bar in the former
than in the latter, and that as a result, bars in exponential
disks could be destructed more easily.  However, \authorcite{ald05}
(\yearcite{ald05}, hereafter ALD) have shown that complete bar
destruction needs a BH with a mass of at least 5\% of the disk
mass even for more realistic models that are composed of an
exponential disk and its surrounding live halo.  Their result
suggests that the difference in the minimum BH mass does not
necessarily originate from the difference in the mass profile
of the disks.  Consequently, the discrepancy in the minimum BH
mass still remains unexplained, although the difference in disk
geometry and situation between the two studies might be related
to that difference in the BH mass: HH used a two-dimensional
disk without a halo, while ALD adopted a three-dimensional
disk with a live halo.

If the BH mass required for destroying a bar is at least about
5\% of the disk mass, bar dissolution of the kind stated here
would be practically impossible.  This is because the corresponding
BH mass amounts to approximately $10^{9.5} M_\odot$ for a typical
disk galaxy, while the largest BH mass derived observationally
from nearby spiral galaxies is about $10^9 M_\odot$
\citep{kormendy88,koretal96,mh03}.  On the other hand,
if that mass is around 0.5\% of the disk mass, we can expect
the existence of bar-dissolved galaxies in the real Universe.
In this case, as \citet{larson10} has argued, dissolving a bar
could weaken and ultimately cease the central activities of
disk galaxies, since a barred structure is considered effective
in fueling gas into the nucleus and this gas inflow could drive
such activities.  Therefore, also from this viewpoint, it is
significant to determine the precise lower limit of the BH mass
for bar destruction in order to reveal the secular evolution of
the barred galaxies that harbor central BHs.  However, it is
conceivable that the BH mass needed to destroy a bar depends
on the characteristics of the bar.  Indeed, the bars formed by
the bar instability will have a diversity of shapes and structures
that arise from the difference in the kinematic structure of the
disk like that represented by \citet{toomre64} $Q$ profile, even
though its mass profile is fixed.  Nevertheless, in the previous
$N$-body studies noted above, the kinematic structure of the disk
for a given mass profile was never changed systematically, with
other parameters of the system intact.  Instead, some parameters
included in the system such as a BH mass and a BH softening length
were surveyed systematically for a disk with given mass and velocity
profiles.

In this paper, we examine what type of bar is destroyed easily
by a massive central BH, and thereby aim at understanding what
causes that difference in the minimum BH mass required for bar
dissolution which has been described above.  In section 2, we
show disk models with different kinematic structures for given
mass profiles in order to produce a wide variety of bars.  In
addition, a BH model is also provided.  Our numerical method is
described in section 3.  In section 4, we present the evolution
of the disk models, in which bars form spontaneously and then
they are deformed by adding a BH at the center of the disk.
The properties of the bars are disclosed as well.  In section
5, we discuss destructible bars under the dynamical influence
of a massive central BH.  Conclusions are given in section 6.

\section{Models}
\begin{figure*}
\centerline{\FigureFile(140mm,40mm){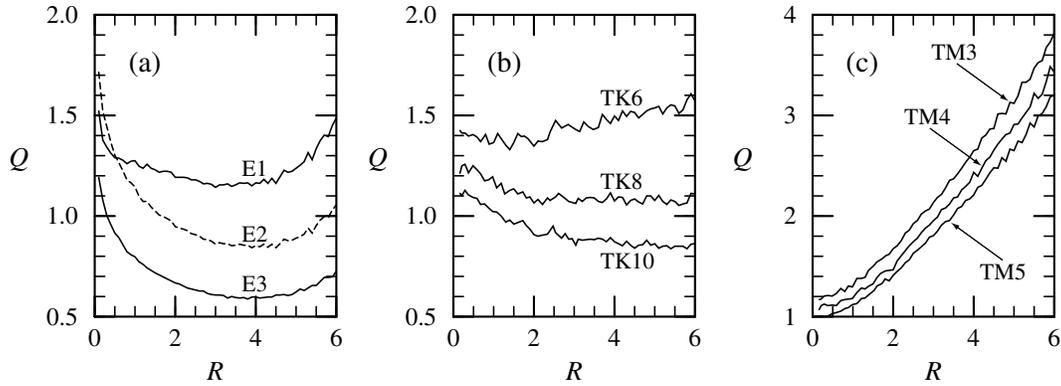}}
\caption{Toomre's $Q$ parameters as a function of radius
for (a) models E1, E2, and E3 of the exponential disks,
(b) models TK6, TK8, and TK10 of the Kuzmin-Toomre disks
with Kalnajs's distribution functions, and (c) models TM3,
TM4, and TM5 of the Kuzmin-Toomre disks with Miyamoto's
distribution functions.  Note that $Q$ values at small radii
for model E1 are smaller than those for model E2.  Here, the
radius is normalized by the exponential scale length $h$.}
\label{fig:Q_profile}
\end{figure*}

We adopt infinitesimally thin stellar disks in order to
avoid another damaging influence on a bar arising from
the fire-hose instability inherent in three-dimensional
disks \citep{raha91,debattista04,mvs04}.  In addition,
we do not include bulge and halo components.  Two
different mass profiles for such flat disks are employed,
that is, exponential and Kuzmin-Toomre disks that are
globally unstable to the formation of a bar.  On the
other hand, as an external field, we model a BH that
is placed at the center of the disk after the full
growth of a bar.

\subsection{Exponential Disks}
The surface-density distributions of exponential disks,
$\mu_{\rm exp}$, are given by
\begin{equation}
\mu_{\rm exp}(R)=\frac{M_d}{2\pi h^2}\exp(-R/h),
\end{equation}
where $M_d$ is the mass of the disk, $h$ is the scale
length, and $R$ is the distance from the center of the
disk.  The disks are truncated at $R=10h$.  As adopted
by HH, the full phase space is realized with the method
devised by \citet{hernquist93}, in which the velocity
structure is approximated using moments of the collisionless
Boltzmann equation.  In so doing, following the observations
of disk galaxies \citep{vdks81,lf89}, we determine the square
of the radial velocity dispersion, ${\sigma_R}^2$, so as to be
proportional to the surface density such that
\begin{equation}
{\sigma_R}^2(R)={\sigma_0}^2\exp(-R/h),
\label{eq:radial_disp}
\end{equation}
where $\sigma_0$ is the radial velocity dispersion at the center.
Although the initial models thus constructed are not based on
exact equilibrium distribution functions, we did not see any
violent changes in the early stages of evolution.  This behavior
indicates that these models are sufficiently close to exact
dynamical equilibrium.

We set up three models that have different \citet{toomre64}
$Q$ profiles shown in figure \ref{fig:Q_profile}a.  These
models are named models E1, E2, and E3, whose reference $Q$
values at the supposed solar radius, 2.43$h$ (8.5 kpc, see
below), are 1.2, 0.91, and 0.64, respectively.  As explained
by \citet{hernquist93}, there is difficulty in completely
fulfilling the ${\sigma_R}^2$ distribution given by equation
(\ref{eq:radial_disp}) near the center, so that some reduction
in ${\sigma_R}^2$ is needed within a certain radius.  As a
result, the ${\sigma_R}^2$ profiles of the models created
here are not shifted vertically to one another at small radii.
This behavior applies also to the $Q$ profiles of the models,
taking into consideration the definition of the $Q$ parameter.
Accordingly, model E1 has a rather different $Q$ profile at
small radii than models E2 and E3.  Thus, in this disk model
group, except at small radii, the functional form of the
velocity structure is the same, and the radial velocity
dispersion at each radius is different from model to model
by a constant factor.

\subsection{Kuzmin-Toomre Disks}
The surface-density distributions of Kuzmin-Toomre (K-T)
disks, $\mu_{\rm KT}$, are provided by
\begin{equation}
\mu_{\rm KT}(R)=\frac{M_d}{2\pi h^2}
  \left(1+\frac{R^2}{h^2}\right)^{-3/2}.
\label{eq:sd_disk}
\end{equation}
The disks are again truncated at $R=10h$.

In contrast to the exponential disks, the K-T disks are
known to possess exact equilibrium distribution functions
(DFs), $F_0$, in analytical forms.  In general, for a flat
axisymmetric galaxy, the equilibrium DF of directly rotating
stars, $F^+$, is composed of the energy, $\varepsilon$,
and angular momentum, $j$, of a star per unit mass.  This
prograde part of the equilibrium DF, $F^+(\varepsilon, j)$,
has been given by \citet{kalnajs76} and by \citet{miya71}.
Here, we use both of the DFs.

On the other hand, there is no definite way of prescribing
the distribution of retrograde stars.  Then, we introduce
them in the same manner as that adopted by \citet{nishida84}.
Consequently, the equilibrium DF is written as
\begin{equation}
F_0(\varepsilon,j)=\left\{
  \begin{array}{ll}
    (1/2)F_0^+(\varepsilon)+F_1^+(\varepsilon,j) & j\ge 0,\\
    (1/2)F_0^+(\varepsilon) & j<0,
  \end{array}\right.
\label{DF}
\end{equation}
where the functions $F_0^+(\varepsilon)$ and $F_1^+(\varepsilon,j)$
are derived from the expansion of $F^+(\varepsilon,j)$ as
\begin{equation}
F^+(\varepsilon,j)=F_0^+(\varepsilon)+F_1^+(\varepsilon,j).
\end{equation}

The two types of DF described above involve a model parameter that
assigns the kinematic structure to the disk.  When denoting this
parameter as $m_{\rm K}$ for Kalnajs's DFs and as $m_{\rm M}$ for
Miyamoto's DFs, we take $m_{\rm K}=6, 8,$ and 10, which are termed
models TK6, TK8, and TK10, respectively, and $m_{\rm M}=3, 4,$
and 5, which are named models TM3, TM4, and TM5, respectively.
We present the $Q$ profiles for models TK6, TK8, and TK10 in
figure \ref{fig:Q_profile}b, and those for models TM3, TM4, and
TM5 in figure \ref{fig:Q_profile}c.  The models generated from
Kalnajs's DFs have slightly declining $Q$ distributions with
radius except for model TK6 that shows a slight increase in
$Q$ from $R\sim 1.3h$ with radius, while those from Miyamoto's
DFs yield steeply rising $Q$ distributions with radius.  At any
rate, as the model parameters, $m_{\rm K}$ and $m_{\rm M}$,
increase, the $Q$ values at all radii decrease for both DFs.

In these disk models, the functional form of the velocity
structure is completely different between these two types
of DF.  For Miyamoto's DFs, the square of the radial velocity
dispersion is represented by
\begin{equation}
{\sigma_R}^2(R)=-\frac{\Phi_{\rm KT}(R)}{2m_{\rm M}+4},
\end{equation}
where $\Phi_{\rm KT}$ is the potential of the K-T disk.
Thus, in this disk model group, the radial velocity dispersion
at each radius is different from model to model by a constant
factor that depends on the model parameter $m_{\rm M}$, so
that the $Q$ profiles are shifted vertically to one another,
as is obvious in figure \ref{fig:Q_profile}c.  On the other
hand, Kalnajs's DFs lead to rather complicated functional forms
of the radial velocity dispersion.  Even so, the resulting $Q$
profiles show sequential changes designated by the model parameter
$m_{\rm K}$, as seen from figure \ref{fig:Q_profile}b.

\subsection{Black Hole}\label{subsec:BH}
A black hole (BH) model is the same as that employed by HH.
That is, a BH is represented by a softened point-mass using
a spline-kernel \citep{hk89}, and its scale length is set to
be $0.01h$.

The BH is added at time $t_{\rm BH}$ after a bar is fully
developed.  To avoid a sudden change in the dynamical state
of the disk caused by an addition of the BH, we increase the
BH mass, $M_\bullet(t)$, from 0 to $M_{\rm BH}$ gradually with
time $t$ as follows:
\begin{equation}
M_\bullet(t)=\left\{
  \begin{array}{ll}
    M_{\rm BH}\tau^2(3-2\tau) & 0 \le \tau \le 1,\\
    M_{\rm BH} & \tau > 1,
  \end{array}\right.
  \label{bhmass}
\end{equation}
where $\tau=(t-t_{\rm BH})/t_{\rm grow}$, and $t_{\rm grow}$
is the time for the BH to attain to its preassigned mass
$M_{\rm BH}$.  Thus, the BH is made to grow adiabatically
by taking $t_{\rm grow}$ to be sufficiently long.  In our
dimensionless system of units described in the next section,
we choose $t_{\rm grow}$ to be 100, and set $t_{\rm BH}$ to
be 200.

According to recent observations, the Sa galaxy NGC 4594
\citep{kormendy88,mh03} and the S0 galaxy NGC 3115 \citep{koretal96}
are thought to harbor a central BH with a mass of approximately
$10^9 M_\odot$, which is the largest BH mass inferred in nearby
spirals.  Since this BH mass is close to 1\% of the disk mass for
a typical disk galaxy, we set $M_{\rm BH}=0.01 M_d$.  With this
choice, we can evaluate the largest impact of a central BH on a
bar from a realistic point of view.

\section{Method}
We use a self-consistent field (SCF) method (Hernquist \& Ostriker
\yearcite{ho92}) to follow the evolution of the disks constructed
in the previous section and that of the systems including the BH.
In this approach, Poisson's equation is solved by expanding the
surface density, $\mu$, and potential, $\Phi$, of the disk in a
bi-orthogonal basis set such that
\begin{equation}
  \mu(\mbox{\boldmath $R$})=
    \sum_{nm} A_{nm}(t)\mu_{nm}(\mbox{\boldmath $R$})
  \label{muexpand}
\end{equation}
and
\begin{equation}
  \Phi(\mbox{\boldmath $R$})=
  \sum_{nm} A_{nm}(t)\Phi_{nm}(\mbox{\boldmath $R$}),
  \label{phiexpand}
\end{equation}
where $\mu_{nm}$ and $\Phi_{nm}$ are, respectively, the
surface-density and potential basis functions, $A_{nm}$
are the expansion coefficients, and $\mbox{\boldmath $R$}$
is the position vector.  In the SCF code, we adopt \citet{ai78}
basis set that is suitable for flat stellar disks.  The outline
of SCF simulations is explained by \citet{ho92} and by
\citet{hozumi97}.  In addition, the computational details
concerning the simulations presented here are described by HH.

The disks are realized with $N=1,000,000$ particles of equal mass.
The equations of motion are integrated in Cartesian coordinates
with a time-centered leapfrog algorithm (e.g., \cite{press86}).
We present results in the system of units such that $G=M_d=h=1$,
where $G$ is the gravitational constant.  These units can be scaled
to physical values appropriate for the Milky Way using $h=3.5$ kpc
and $M_d=5.6\times 10^{10} M_\odot$, with the result that unit time
and velocity are $1.31\times 10^7$ yr and 262 km s$^{-1}$,
respectively.

We set the maximum numbers of radial and azimuthal expansion
terms in the SCF code, $n_{\rm max}$ and $m_{\rm max}$,
respectively, to be $n_{\rm max}=24$ and $m_{\rm max}=12$.
In imposing these numbers of the expansion terms on each model,
we save only even $m$-values to avoid a sizable disagreement
between the center of the bar and the position of the BH
caused by asymmetric features that would originate from
odd $m$-values.  Again, the computational details are
taken over from those stated by HH.

Although our present simulations include a wide range of timescales
after introducing the BH, we use a fixed time step that is made
as small as possible.  In order to reduce computation times, we
have parallelized the SCF code in accordance with the prescription
provided by \citet{hsb95} (see also \cite{kelly01}).  The time
steps adopted are $\Delta t=0.05$ and $\Delta t=0.002$ before and
after the addition of the BH, respectively.  With these choices of
time step, the relative energy error of the system after the full
growth of the BH was, in all cases, smaller than $7.2\times 10^{-6}$.

\section{Results}

\subsection{Evolution of Bars}
\begin{figure}
\centerline{\FigureFile(70mm,150mm){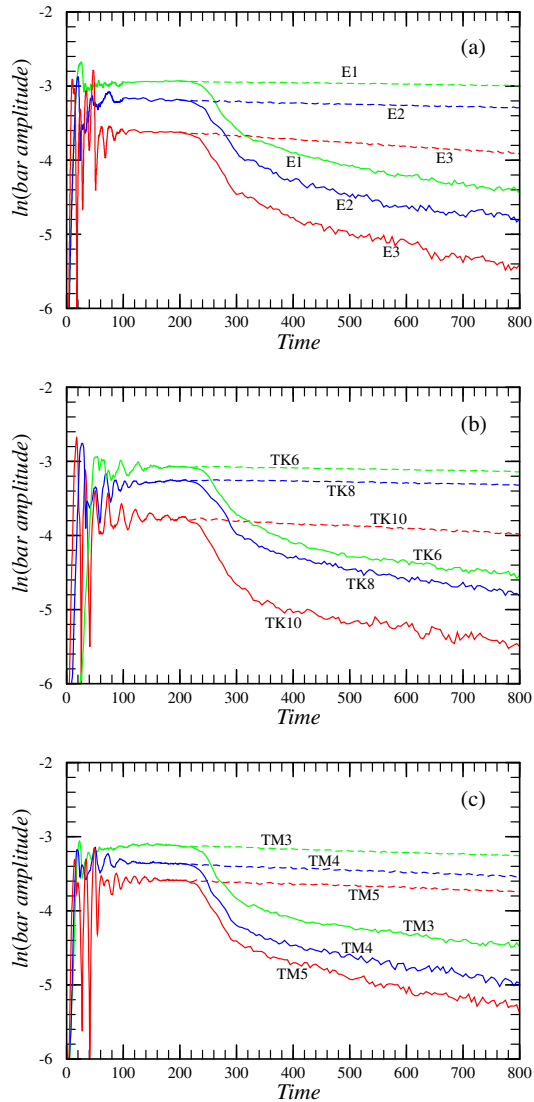}}
\caption{Time evolution of the bar amplitude before and after
the addition of the black hole (solid lines) for (a) models E1,
E2, and E3 of the exponential disks, (b) models TK6, TK8, and
Tk10 of the Kuzmin-Toomre disks with Kalnajs's distribution
functions, and (c) models TM3, TM4, and TM5 of the Kuzmin-Toomre
disks with Miyamoto's distribution functions.  The black hole
is added at $t=200$, and grows up fully at $t=300$.  The bar
evolution without the black hole after $t=200$ is also shown
with dashed lines for each model.}
\label{fig:baramp}
\end{figure}

\begin{figure*}
\centerline{\FigureFile(150mm,60mm){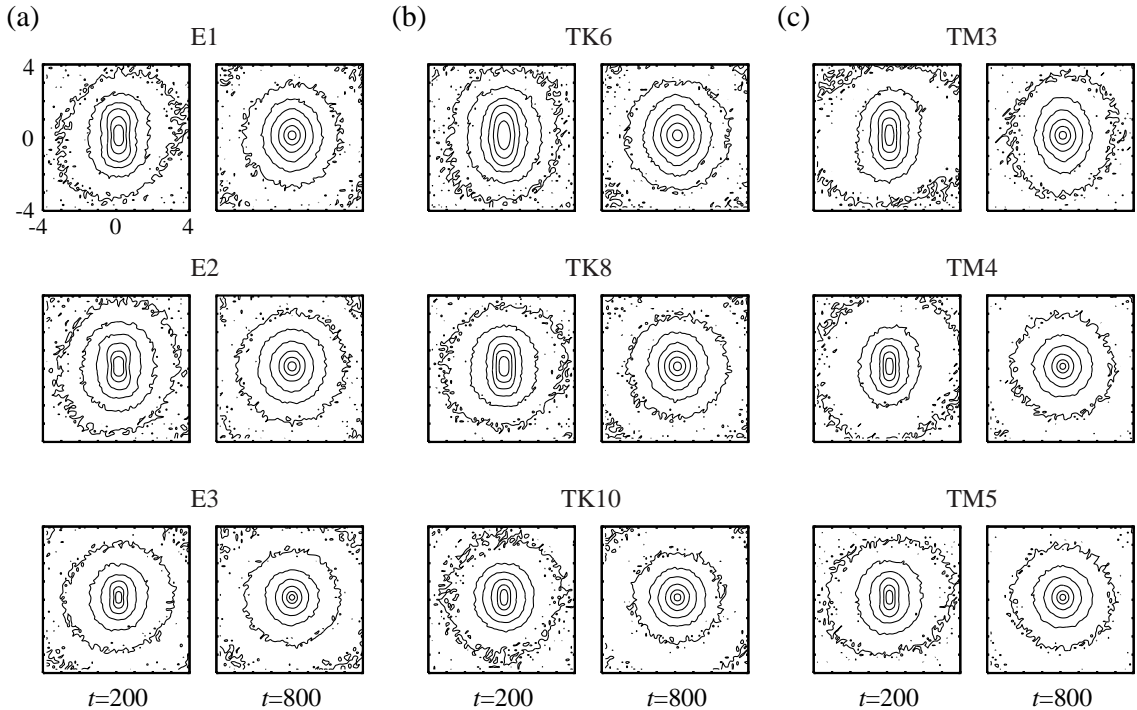}}
\caption{Density contours in barred structures at the time
just before the addition of the black hole ($t=200$) and at
the end of the simulations ($t=800$) for (a) models E1, E2,
and E3 of the exponential disks, (b) models TK6, TK8, and
TK10 of the Kuzmin-Toomre disks with Kalnajs's distribution
functions, and (c) models TM3, TM4, and TM5 of the Kuzmin-Toomre
disks with Miyamoto's distribution functions.  The bar major
axis is aligned to the $y$ axis.  The contour levels are drawn
at the 90, 80, 70, $\cdots$, and 30\% of the peak amplitude on
logarithmic scales.  Black hole growth starts at $t=200$ and
ends at $t=300$.  All patterns rotate counterclockwise.}
\label{fig:contour}
\end{figure*}

As the disks are evolved forward in time, linearly growing
two-armed spiral modes excited in the disks are deformed
into bars in nonlinear regimes through the bar instability.
We evaluate the bar amplitude by the absolute value of the
expansion coefficient, $|A_{22}(t)|$, which corresponds to
the amplitude of the fastest growing two-armed mode (see HH).
Figure \ref{fig:baramp} shows the time evolution of the bar
amplitude for each model of the three disk model groups.
From this figure, we find that all bars are fully developed
at least by $t=200$, and that their amplitudes are practically
held constant over time to the end of the simulations unless
the BH is added, although in an exact sense, some bars grow
in amplitude very slightly with time and others show small
decrements in amplitude with time, in response to a way of
redistributing the angular momentum of a bar.  We have thus
chosen the time of adding the BH to be $t_{\rm BH}=200$, as
mentioned in section \ref{subsec:BH}.  We notice from figure
\ref{fig:baramp}, along with figure \ref{fig:Q_profile}, that
in each disk model group, the bar amplitude becomes lower
as the disk is colder in the sense of typical $Q$ values.
Concerning the exponential disks, the bar amplitude of model
E1 is larger than that of model E2, while $Q$ values at
small radii for model E1 are slightly smaller than those for
model E2 as expressed in figure \ref{fig:Q_profile}.  This fact
implies that the bar amplitude would not be sensitive to the
inner velocity structures but to the relatively outer ones.

\begin{figure}
\centerline{\FigureFile(70mm,150mm){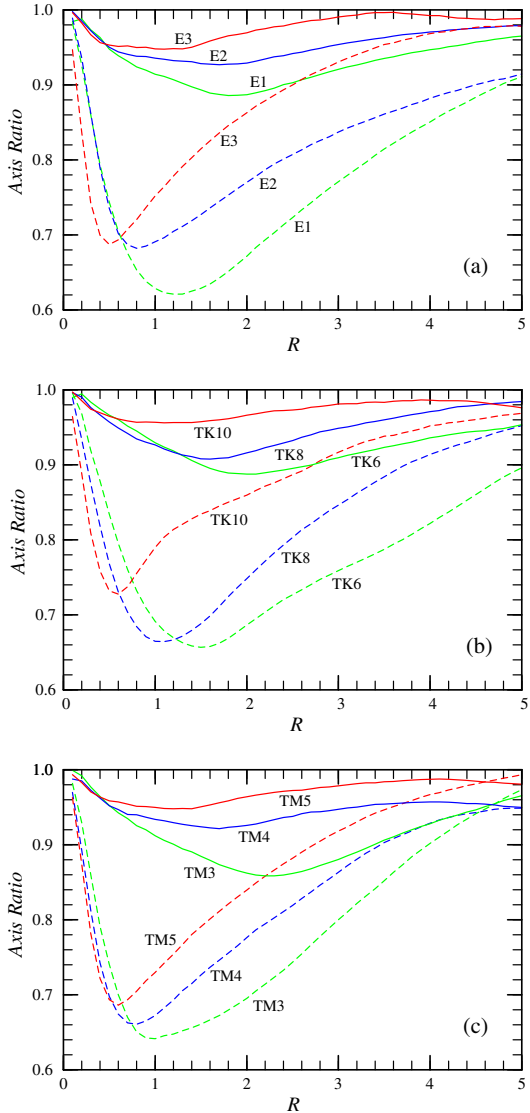}}
\caption{Axis ratio of bars as a function of radius for (a) the
exponential disks, (b) the Kuzmin-Toomre disks with Kalnajs's
distribution functions, and (c) the Kuzmin-Toomre disks with
Miyamoto's distribution functions.  Dashed lines show the axis
ratios of the bars just before the black hole is added ($t=200$),
while solid lines represent those at the end of the simulations
($t=800$).}
\label{fig:axisratio}
\end{figure}

Once the BH is added at $t=t_{\rm BH}$, the bar amplitudes start
to decrease with time, as demonstrated in figure \ref{fig:baramp}.
After the full growth of the BH, the bar amplitudes continue
to decay almost exponentially with time till the end of the
simulations.  Consequently, the bars become very round in the
final stages as found from a comparison between the density
contours at $t=t_{\rm BH}$ and those at the end of the
simulations, which are illustrated in figure \ref{fig:contour}.
To quantify to what degree the bars are made round, we evaluate
the axis ratios of the bars along radius at these two times
by calculating the principal moment of inertia tensor for
particles included in a specified radius, the value of which
is regarded as the axis ratio at that radius.  Resulting
axis-ratio profiles are presented in figure \ref{fig:axisratio}.
We can see from this figure, together with figure \ref{fig:contour},
that in each disk model group, the bar is more easily deformed
into a rounder shape as the smallest axis ratio along radius is
larger, that is, the bar is rounder when the BH is added, or as
the bar half-length, estimated here to be a distance from the
center to the radius of the smallest axis ratio, is shorter.

Since the final axis ratios of the bars exceed about 0.9 at all
radii except for the bar of model TM3 (the final axis ratio is
0.86), irrespective of the mass and velocity profiles, it follows
that a BH with a mass of 1\% of the disk mass can destroy almost
completely, or in some cases appreciably, the bars formed by the
bar instability.  Furthermore, when we take into account the $Q$
profiles shown in figure \ref{fig:Q_profile}, the runs of the final
axis ratios in figure \ref{fig:axisratio} indicate that in each
disk model group, the bars originating from the colder disks
are destroyed more easily by the BH.

\subsection{Characteristics of Bars}
We describe the characteristics of the bars before the BH acts
on them to unravel what type of bar is more destructible under
the dynamical influence of a massive central BH.

\begin{figure*}
\centerline{\FigureFile(140mm,70mm){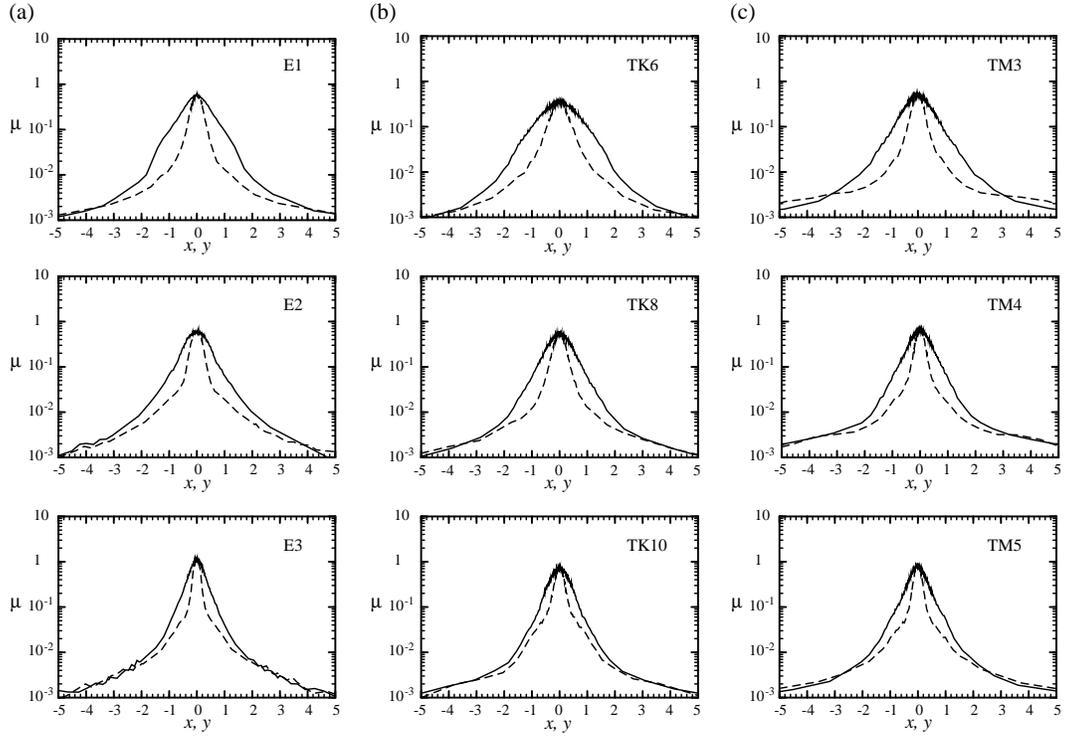}}
\caption{Surface-density profiles along the major (solid lines)
and minor (dashed lines) axes at the time just before the black
hole is added ($t=200$) for (a) models E1, E2, and E3 of the
exponential disks, (b) models TK6, TK8, and TK10 of the
Kuzmin-Toomre disks with Kalnajs's distribution functions,
and (c) models TM3, TM4, and TM5 of the Kuzmin-Toomre disks
with Miyamoto's distribution functions.}
\label{fig:sd}
\end{figure*}

Figure \ref{fig:sd} presents the surface-density profiles
of the bars along the major and minor axes at $t=t_{\rm BH}$
just before the BH is added.  Except for models E1 and
TK6, there is no noticeable change in the density slope
on logarithmic scales from the center towards the bar
ends along the major axis.  On the other hand, models E1
and TK6 show a bend in the surface-density distribution
along the major axis at around an intermediate position
between the center and the bar end, although the bend for
model TK6 is rather gentle as compared to that for model E1.
The surface-density profiles along the minor axis for these
two models are close to those of the rest of the models.

\begin{figure*}
\centerline{\FigureFile(140mm,70mm){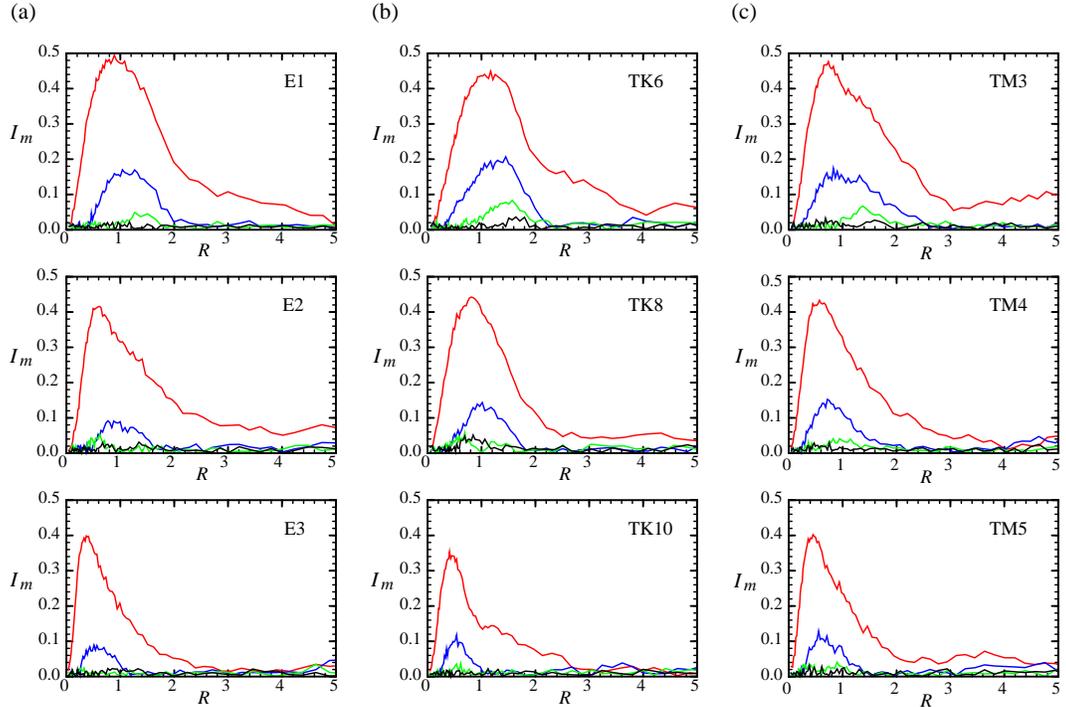}}
\caption{Fourier amplitudes of the surface-density distribution,
$I_m$, for the $m=2$ (red lines), 4 (blue lines), 6 (green lines),
and 8 (black lines) components normalized by the $m=0$ component
at the time just before the black hole is added ($t=200$) as a
function of radius for (a) models E1, E2, and E3 of the exponential
disks, (b) models TK6, TK8, and TK10 of the Kuzmin-Toomre disks
with Kalnajs's distribution functions, and (c) models TM3, TM4,
and TM5 of Kuzmin-Toomre disks with Miyamoto's distribution
functions.}
\label{fig:fourier}
\end{figure*}

To examine the structures of the bars in detail, we analyze
the surface-density profiles by calculating their Fourier
components along radius, $I_m(R)$, where $m=2, 4, 6,$ and 8.
These Fourier components are normalized by the $m=0$ component,
$I_0$, of the disk being analyzed.  The resulting relative
Fourier amplitudes of each model at $t=t_{\rm BH}$ are displayed
in figure \ref{fig:fourier}.  Here, we calculate the Fourier
components in the same manner as that done by \citet{ohw90}
and by \citet{am02}.  This figure indicates that for the hottest
models in each disk model group from a viewpoint of typical $Q$
values, that is, for models E1, TK6, and TM3, the amplitudes for
the $m=4$ and $m=6$ components are large in magnitude as well as
for the $m=2$ component, while for the rest of the models they
are small, and, in particular, the amplitudes for the $m=6$
component are negligibly small.  In addition to this property
of the hottest models, taking into account the $Q$ profiles
presented in figure \ref{fig:Q_profile}, together with figure
\ref{fig:sd}, we can see that a bend in the surface-density
distribution of the bar along the major axis is produced for
the models that have relatively high $Q$ values near the center
as for models E1 and TK6.

To further characterize the bars, we measure the size of the bar
region at $t=t_{\rm BH}$, following Ohta \etal\ (\yearcite{ohw90}),
who defined the bar region to be the region where the value of
$(I_0+I_2+I_4+I_6)/(I_0-I_2+I_4-I_6)$ exceeds 2.0.  In figure
\ref{fig:barsize}, the measured size of the bar region (bar
half-size) is plotted against the bar amplitude that is
normalized by the amplitude of the ring mode ($m=0$),
$|A_{22}|/|A_{00}|$, at $t=t_{\rm BH}$ in order to be compared
among the disk models with the different mass and velocity profiles.
This figure reveals that the bar size is approximately proportional
to the normalized bar amplitude.

As another ingredient for featuring the bars, the smallest axis
ratio along radius at $t=t_{\rm BH}$, as a measure of the degree
of elongation, is plotted against the normalized bar amplitude
in figure \ref{fig:barshape}.  We can find from this figure that
on the whole, the roundness of the bar decreases monotonically
with increasing normalized bar amplitude.

\section{Discussion}
\begin{figure}
\centerline{\FigureFile(70mm,50mm){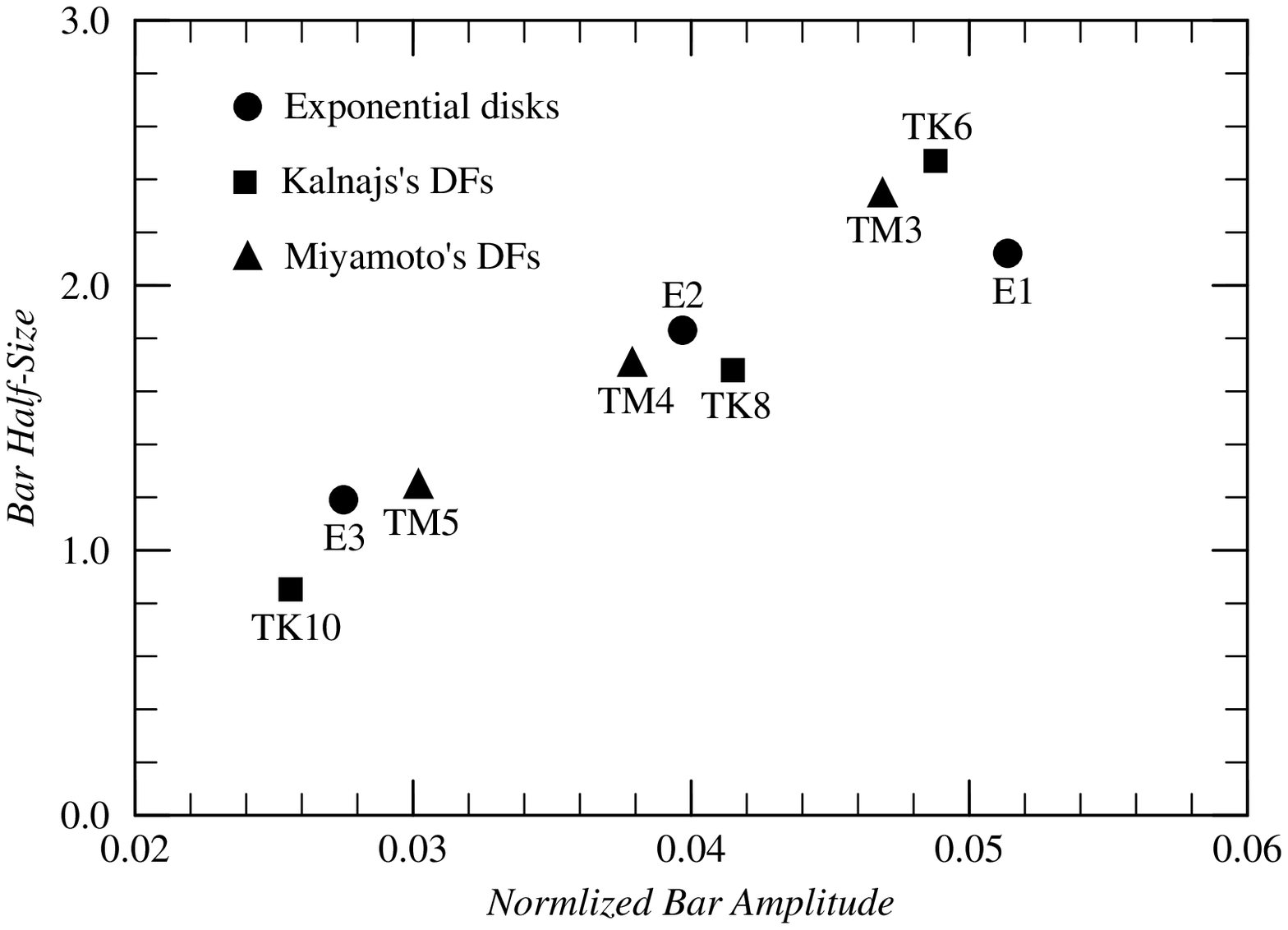}}
\caption{Bar half-size as a function of bar amplitude normalized
by the amplitude of the ring mode excited in each model disk.
The bar half-size and the amplitudes of the bar and ring modes
are calculated at the time just before the addition of the black
($t=200$).  The definition of the bar half-size is described
in the text.}
\label{fig:barsize}
\end{figure}

\begin{figure}
\centerline{\FigureFile(71.186mm,53.5mm){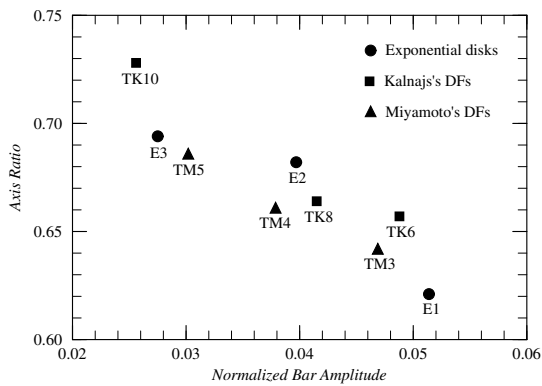}}
\caption{Axis ratio of bars as a function of bar amplitude
normalized by the amplitude of the ring mode excited in each
model disk.  The axis ratio means the smallest axis ratio along
radius for each model.  The axis ratio and the amplitudes of
the bar and ring modes are calculated at the time just before
the addition of the black hole ($t=200$).}
\label{fig:barshape}
\end{figure}

We have obtained a wide variety of bars from the exponential
and K-T disks to which various velocity distributions are
assigned.  The amplitude of these bars continues to decay
exponentially with time till the end of the simulations
after the full growth of the BH that is added at the center
of the disk.  When we pay attention to the three models of
each disk model group, the final axis-ratio profiles in
figure \ref{fig:axisratio} show that the BH can make the
bar rounder as the bar amplitude becomes lower.  To be sure,
it may be no wonder that lower-amplitude bars are destructed
more easily by a given mass BH.  The important thing is that
this fragility of low-amplitude bars originates from the
bar properties exposed in figures \ref{fig:barsize} and
\ref{fig:barshape} that the normalized bar amplitude is
related to the size and shape of a bar.  Therefore, the
precise explanation is that since lower-amplitude bars are
shorter in length and rounder in shape, a given mass BH can
provide a more destructive impact on shorter bars that lie
more deeply inside the BH's ``sphere of influence", and can
increase the roundness more easily for rounder bars over a
given period.

We should mention the bar properties extracted in the present
study from an observational point of view.  \citet{elmegreen07}
have found from $K_s$-band images of barred galaxies that longer
bars have higher bar amplitudes.  On the other hand, analyzing
near-infrared images of nearby barred galaxies, \citet{menendez07}
have indicated a weak tendency that the bar ellipticity is larger
for longer bars.  Here, the bar ellipticity is defined as $1-b/a$,
where $b/a$ is the axis ratio of a bar.  Also, very recently,
\citet{hoyle11} have presented a similar correlation between the
bar ellipticity and bar length from an analysis of bar length
measurements in local disk galaxies.  Then, if we are allowed to
combine these two kinds of observed results, we are led to the
conclusion that the bar ellipticity increases with increasing bar
amplitude.  In fact, the bar ellipticity is often used as a measure
of the bar strength: larger bar ellipticities mean higher bar
amplitudes \citep{chapelon99, laine02, laurikainen02}.  Therefore,
the properties that we have uncovered for the simulated bars are
consistent with those of real barred galaxies.  Although the
correlation of the bar length and ellipticity with the bar
amplitude has thus far been explained from a viewpoint
of secular evolution \citep{elmegreen07, menendez07, hoyle11},
we will demonstrate in a future paper that these bar properties
are closely related to the properties of the fastest growing,
globally unstable two-armed modes excited in stellar disks,
which finally lead to bars in nonlinear regimes.

We have also found that the lower-amplitude bars are generated
from the colder disks in each disk model group, although the
physical origin of this finding is unclear at the moment (we
will mention it also in that future paper).  Consequently,
once the functional form of the velocity distribution is
assigned, the degree of the coldness of a disk determines
the fragility of a bar under the dynamical influence of a
massive central BH at least for two-dimensional disks.  If
this property is applicable to bars formed in three-dimensional
disks, the discrepancy that HH's minimum BH mass necessary
for bar dissolution in the exponential disk is about an
order of magnitude smaller than ALD's one might have arisen
partly from the difference in the coldness, more precisely,
the $Q$ profile of the disk.  However, it is uncertain whether
the relation between the coldness of a disk and the amplitude
of a bar holds for bars produced in three-dimensional disks,
because the fire-hose instability intrinsic in such disks
\citep{raha91,debattista04,mvs04} and bar-halo interactions
(\cite{lia02}; Athanassoula \& Misiriotis \yearcite{am02}),
both of  which are not included in the present study, might
distort those properties of the bars which have been obtained
in flat disks.  Thus, we need to investigate the validity of this
relation using realistic disk galaxy models in three-dimensional
configurations.

Note that the relation between the coldness of a disk and the
fragility of a bar mentioned above is obtained by ignoring the
effects of gas components on the bar instability, because our
simulations are based on pure stellar dynamics.  In general,
if gas components are added to a stellar disk, the system
will become colder on account of their low velocity dispersions.
Then, we could expect a lower-amplitude bar in such a system
that would suffer a more violent bar instability.  In a case
where a large gas mass fraction is included in a disk, \citet{sn93}
demonstrated that the bar instability in the disk is damped heavily.
Even in this situation, it can be interpreted apparently that an
extremely weak bar or bar-like structure is formed in an exceedingly
cold disk owing to a large amount of gas.  Thus, in principle, our
finding that colder disks lead to lower-amplitude bars might hold
even for the disks that contain gas components, regardless of
their amount.

Figures \ref{fig:sd} and \ref{fig:fourier} have indicated
that robust bars against the dynamical influence of a massive
central BH have relatively larger Fourier amplitudes for the
$m=4$ and $m=6$ components (models E1, TK6, and TM3), and that
for the models whose $Q$ values are somewhat large near the
center (models E1 and TK6), the bars represented by these Fourier
components exhibit a bend in the surface-density profile along
the major axis.  These two features appear similar to those of
the MH-type models, employed by \citet{lia02} and by \citet{am02},
whose halos are characterized by small core radii.  On the other
hand, the surface-density profiles of the models other than models
E1 and TK6 have no noticeable change in density slope along the bar
major axis.  This feature corresponds to that obtained from the
MD-type models, also used by \citet{lia02} and by \citet{am02},
whose halos are depicted by large core radii.  ALD have reported
that for a given mass central BH, bars formed in the MH-type
models are less destructible than those in the MD-type models.
Interestingly, ALD mentioned that the disks in the MH-type models
have $Q=1.4$, while those in the MD-type models have $Q=1$.  Thus,
this relation between the degree of bar destruction and the typical
$Q$ value is in concord with our claim that bars formed in colder
disks are more destructible.  It is true that we cannot compare
directly between their results and ours, since the disks in their
MH- and MD-type models are three-dimensional and their halos are
self-gravitating, while our models are two-dimensional and have
no halos.  Nevertheless, our models seem to reflect the
characteristics possessed by the bars in their MH- and MD-type
models.  This finding may imply that the velocity structure of
the disk affects more substantially the characteristics of the bars
formed by the bar instability than the difference in the central
concentration of self-gravitating halos.  Therefore, we should
examine the bar properties for both MH- and MD-type models that
have a wide variety of velocity structures in the disks, and
evaluate the minimum BH mass needed to dissolve the bars formed
spontaneously.  This line of investigation will enable us to
confirm whether the discrepancy in the minimum BH mass between
HH's study and the others is due to the difference in the velocity
structure of the disk.

HH argued that the destructive effect of a central BH on a
bar could be considered stronger for more centrally concentrated
disks.  Although Norman \etal\ (1996) used a K-T disk and ALD
adopted an exponential disk, both studies resulted in a similar
BH mass that is about 5\% of the disk mass as the minimum value
required for bar dissolution, in spite of the fact that exponential
disks are more centrally concentrated than K-T disks.  This result
might suggest that the velocity structure is more significant to
bar dissolution than the degree of mass concentration of the disk,
because our simulations have shown that the vulnerability of a bar
to a given mass central BH depends on the velocity structure of the
disk.  Furthermore, we have found from figures \ref{fig:barsize}
and \ref{fig:barshape} that the size and roundness of a bar
are determined by the normalized bar amplitude, irrespective
of the mass profile of the disk, and that the strength of this
amplitude is connected to the destructibility of a bar.  However,
we cannot answer exactly whether these bar properties
depend on the degree of mass concentration of the disk, since
it is prohibitively difficult to construct such models that
have the same velocity structure for different mass profiles.
Therefore, we cannot conclude definitely that the bars generated
in more centrally concentrated disks are destroyed more easily
by a given mass central BH.

We have found that the bar amplitude continues to decrease
almost exponentially with time for 500 time units after the
full growth of the BH.  This simulated period corresponds to
about 7 Gyr, if the time unit is represented by the physical
value appropriate for the Milky Way.  We have also found that
the bars formed in the exponential disks have been practically
destroyed within that period by the BH whose mass is 1\%
of the disk mass.  This BH mass is reduced to be approximately
$10^9 M_\odot$ for a typical disk galaxy, and so, it is
comparable to the largest mass for central BHs in nearby
spirals \citep{kormendy88,koretal96,mh03}.  Therefore,
our simulations suggest that bar dissolution induced by
a massive central BH could occur in the real Universe, if
galaxy disks are relatively cold at birth.  However, the
frequency of such bar dissolution would be rare, since the
minimum BH mass examined by HH (about $10^{8.5} M_\odot$)
as well as the BH mass in this work are close to the upper
bound of the observationally derived BH masses.  Consequently,
the bar fraction would be declined to a negligible degree,
if any, as the Universe evolves, by the bar dissolution
mechanism studied here.  In fact, \citet{elmegreen04}
and \citet{jogee04} found that the observed bar fraction
is nearly constant, and recently, \citet{sheth08} have
shown that it even increases with time over the last 7
Gyr (see also \cite{abraham99,vdB02}).

\section{Conclusions}
We have examined the properties of the bars formed by the bar
instability in the exponential disks with approximate equilibrium
velocity distributions and the Kuzmin-Toomre disks with two
different types of exact equilibrium DF in two-dimensional
configurations, and the damaging impact of a massive central
BH on these bars.  Then, we have found that lower-amplitude
bars, that is, weaker bars are destroyed more easily by a given
mass central BH, because the bar formed by the bar instability
is shorter in length and rounder in shape as the bar amplitude
decreases, thereby being more susceptible to the dynamical
influence of that BH.

In addition, we have also found that lower-amplitude bars are
generated from colder disks for a given functional form of the
velocity distribution.  This finding suggests that destructible
bars under the dynamical influence of a given mass central BH
are determined by the kinematic properties of disks, like those
specified by typical $Q$ values, at the formation epoch, once
the functional form of the velocity distribution is assigned.
Thus, it is likely that the discrepancy in the lowest BH mass
required for bar dissolution found in the previous simulations
might have resulted from the difference in the velocity structure
of the disk.

When scaling the dimensionless units to appropriate physical
values, we have shown that the bar amplitude continues to
decrease almost exponentially with time for about 7 Gyr after
the full growth of a BH, and that the bars formed in the relatively
cold exponential disks have resulted in practical dissolution for
a BH with a mass of approximately $10^9 M_\odot$, a likely largest
BH mass derived observationally from nearby spirals.  We thus infer
that bar dissolution induced by a massive central BH could occur in
the real Universe, if galaxy disks are formed in a rather cold state.

\bigskip
The author is indebted to Dr.\ K.\ Nitadori for optimizing the
two-dimensional SCF code for parallelization.  He would also
like to thank the anonymous referee for the valuable comments
that have called the author's attention to the connection of
the present work to observations, which have helped improve
the manuscript.  Numerical computations were carried out on
the PC cluster and Cray XT4 at the Center for Computational
Astrophysics (CfCA), the National Astronomical Observatory
of Japan.

\end{document}